# Tuning Water Slip Behavior in Nanochannels using Self-Assembled Monolayers

Dezhao Huang[1], Teng Zhang[1], Guoping Xiong[2], Linji Xu[3], Zhiguo Qu[4,*], Eungkyu Lee[1,*],

Tengfei Luo[1,5,*]

1. Department of Aerospace and Mechanical Engineering, University of Notre Dame, Notre Dame, Indiana 46556, USA

2. Department of Mechanical Engineering, University of Nevada – Reno, Reno, Nevada 89557, USA

3. Chongqing Research Academe of Environmental Science, Chongqing 401120, China

4. Moe Key Laboratory of Thermo-Fluid Science and Engineering, Energy and Power Engineering School, Xi'an Jiaotong University, Xi'an 710049, China

5. Department of Chemical and Biomolecular Engineering, University of Notre Dame, Notre Dame, Indiana 46556, USA

*Corresponding authors:  Email: tluo@nd.edu, elee18@nd.edu, zgqu@mail.xjtu.edu.cn

**Abstract**

Water slip at solid surfaces is important for a wide range of micro/nano-fluidic applications. While it is known that water slip behavior depends on surface functionalization, how it impacts the molecular level dynamics and mass transport at the interface is still not thoroughly understood. In this paper, we use nonequilibrium molecular dynamics simulations to investigate the slip behavior of water confined between gold surfaces functionalized by self-assembled monolayer (SAM) molecules with different polar functional groups. We observe a positive-to-negative slip transition from hydrophobic to hydrophilic SAM functionalizations, which is found related to the stronger interfacial interaction between water molecules and more hydrophilic SAM molecules. The stronger interaction increases the surface friction and local viscosity, making water slip more difficult. More hydrophilic functionalization also slows down the interfacial water relaxation and leads to more pronounced water trapping inside the SAM layer, which both impede water slip. The results from this work will provide useful insights to the understanding of the water slip at functionalized surfaces and design guidelines for various applications.

## INTRODUCTION

Understanding and controlling the molecular level fluid properties and behavior on solid surfaces is important to a wide range of micro/nano-fluidic applications.[1–5] The interfacial fluidic behavior is even more important at nanoscale channels due to the large surface to volume ratio, where surface effects can play a dominant role in controlling the flow properties.[1–3] There have been extensive studies on liquid slip behavior on solid surfaces. Thompson and Troian first investigated the slip of a Lennard-Jones (L-J) fluid at a solid wall using molecular dynamics (MD) simulations, showing that the slip length increases and begins to diverge as the shear rate approaches a critical value.[6,7] This general nonlinear relationship has later been observed in other non-equilibrium MD (NEMD) simulations.[8–10] The shear rate-dependent slip phenomenon is also reported in surface force experiments in which the slip appears when the critical shear rate is achieved.[11,12] Many researchers have also examined how different parameters like the solid structural factor and nanoscale confinement affect the level of slip at various liquid-solid interfaces.[22,27–33] Using a model system, Priezjev investigated the shear rate dependence of the slip length in a thin polymer film confined between atomically flat surfaces, and he observed an interesting negative-to-positive slip transition when shear rate increases.[22] In another model MD study, Priezjev found altering the polymer melt-wall interaction can lead to changes in the fluid structure near the surface and thus change the slip behavior.[23] Ewen et al.[13] studied realistic n-hexadecane liquids confined between organic friction modifier (OFM)-treated surfaces using NEMD shear simulations to show that OFM can also greatly reduce the friction and the slip length. These studies indicate that modifying surface chemistry can be an effective means to control the liquid slip behavior. Practically, functionalization using self-assembled monolayer (SAM) is a versatile method for surface modification due to their stable covalent bonds with substrates and easily tunable functional groups.[18,24–26] While it is expected that more hydrophilic interface will increase the friction of water and decrease slip, how the molecular level liquid dynamics is influenced and how such influences impact water slip behavior have not been thoroughly studied for water in contact with SAM-functionalized surfaces.

In this study, we use NEMD to study the slip phenomenon of water confined between SAM-functionalized gold (Au) surfaces using realistic interatomic potentials. Three kinds of alkane thiol SAM molecules with different end groups featuring increasing hydrophilicities (-CH$_3$ → -OH → -COOH) are studied to investigate their impacts on the water slip behavior and friction properties. Under the shear rate range ($10^{10}$-$10^{11}$/s) studied, the water slip length for each SAM functionalization is found to be almost constant. However, from hydrophobic SAM to hydrophilic SAM, the slip behavior changes dramatically with the slip length transitioning from positive to negative value. It is found that the stronger interfacial interaction between water molecules and more hydrophilic SAM molecules increase the surface friction and local viscosity, making water slip more difficult. More hydrophilic functionalization also slows down the interfacial water relaxation and leads to more pronounced water trapping inside the SAM layer, which both impede water slip. The results from this work will provide useful insights to the understanding of the water slip at functionalized surfaces and design guidelines for various micro-/nano-fluidic applications.

**METHODS AND SIMULATION MODEL**

In the simulations, the model consists of 4000 water molecules confined between two SAM-functionalized Au slabs as shown in Figure 1a. SAM-functionalized (111) Au slabs with the dimensions of ~ 80 Å ($y$) × 75 Å ($y$) × 22 Å ($z$) are used as the substrates. Three different types of thiol SAM molecules, including 1-Hexanethiol (HS-(CH$_2$)$_5$-CH$_3$), 6-Mercapto-1-hexanol (HS-(CH$_2$)$_6$-OH) and 6-Mercaptohexanoic acid (HS-(CH$_2$)$_5$-COOH) are studied (Fig. 1c). For brevity, they are referred to as −CH$_3$ SAM, −OH SAM, and −COOH SAM, respectively. Periodic boundary conditions are applied in the x- and y-directions. SAM molecules are covalently bound to Au atoms and placed with the spacing of 0.497 nm in a 2D rhombic lattice[27] and the initial tilt angle of 30º. The thickness of the water layer is around 20 Å. It is worth mentioning that the water layer is sufficiently thick so that no apparent surface-induced structuring appears in the middle of the liquid.[28]

Water molecules are modeled using the TIP3P model,[29] which can reproduce water  structural properties well.[30] To ensure that the observed physics is not model-dependent, we have also used the SPC/E[31] water model to perform certain simulations, which are included in the Supporting Information, to confirm that the slip transition from positive to negative slip is still observed. The SAM molecules are modeled using the polymer consistent force field (PCFF),[32] which has been successfully used for interface simulations involving SAMs.[33,34] The non-bond interactions between Au and other atoms are simulated by the L-J interaction:

$$E = 4\varepsilon \left[ (\frac{\sigma}{r_{ij}})^{12} - (\frac{\sigma}{r_{ij}})^6 \right] \qquad (1)$$

where $\varepsilon$ and $\sigma$ are the energy and length constants respectively, and $r_{ij}$ is the distance between two atoms, $i$ and $j$. L-J interaction parameters are documented in our previous works,[33,35] which were modified from the Universal Force Field.[36] A cutoff of 8 Å is used for the Morse potential, and 10 Å is chosen for the L-J interactions. The long-range electrostatic interaction in the entire system is computed by the PPPM (particle-particle particle-mesh) approach with an accuracy of $1 \times 10^{-5}$. Simulations are performed using the large-scale atomic/molecular massively parallel simulator (LAMMPS).[37] The chosen time step size is 1 fs.

First, the system is energy-minimized and equilibrated in a canonical ensemble (NVT) at 300 K for 0.5 ns. Then, the system is optimized in an isothermal-isobaric ensemble (NPT) at 1 atm and 300 K for another 3 ns. After the structures are fully relaxed, the water is sheared by translating the top slab at constant speeds of 100, 150, 200, 250, 300 m/s, which corresponds to shear rates on the order of $10^{10}$-$10^{11}$/s, similar to those used in other NEMD simulations.[9,22,38–41] Once the velocity profile of the confined water reaches steady state, fluid properties are both time- and space-averaged. The criterion to spatially divide the water slabs for velocity averaging is a balance between the statistical noise and resolution.[42] We adopt the rigid Au substrate model, where the Au atoms are frozen, which is common in confined fluid NEMD simulations,[13,22,43,44] but the SAM molecules are allowed to thermally vibrate besides also moving together with the attached Au substrate. The viscous heat generated during the sliding simulations is dissipated using a thermostat acting on the SAM molecules.[13] This indirect heat dissipation method can overcome the disadvantage of directly thermostating the fluid which perturbs liquid molecular dynamics.[8,45]

**RESULTS AND DISCUSSION**

The velocity profiles are computed at different imposed shear velocities of 100, 150, 200, 250 and 300 m/s for the water confined between the two functionalized substrates. Wall velocities are chosen to be sufficiently large to obtain high signal-to-noise ratios in the velocity profile. Figure 1b shows a representative steady-state velocity profile. The slip length is calculated by extrapolating the measured velocity profile to the point at which it intersects the substrate velocity, and the distance between the extrapolated point and the SAM/water interface is designated as the slip length (indicated by the green double arrow in Figure 1b). It should be mentioned that the slip length presented here is the apparent slip length which does not necessarily occur right at the SAM/water interface but may happen at water/water interfaces due to an adsorbed layer of water near the solid-water interface, which happens when the cohesive energy in liquid is stronger than the adhesion energy at the solid-liquid interface.[46]

The extrapolated slip length is shear rate-independent for the three different surface functionalizations as can be seen from Figure 2. For the -CH$_3$ SAM functionalized interface (Figure 2a), due to the weaker wall-water interaction, the extrapolated slip length is 0.90 nm, which is the largest among all three different surface functionalizations. For the more hydrophilic -OH SAM surface (Figure 2b), the slip length decreases to 0.05 nm, and then for the most hydrophilic interface (-COOH SAM, Figure 2c), the slip length transitions further to -0.10 nm. The change from a positive to a negative value of the slip length implies the so-called no-slip boundary shifts from the solid domain to the water domain when the surface is becoming more hydrophilic. One of the main reasons affecting the fluid slip behavior is the wall-fluid interaction strength,[22] and our above observation is a real-case manifestation of this effect. As the wall-fluid interaction strength increases from the hydrophobic -CH$_3$ SAM surface to the hydrophilic -COOH SAM surface, the liquid layer neighboring to the surface becomes more strongly adhered to the wall, and thus the effective no-slip boundary plane moves into the liquid region. At low interface adhesion, the water layer immediately neighboring to the surface can slide relative to the wall when subject to the shear stress from the bulk water. This phenomenon is usually referred to as a molecular slip, and it is the case for the -CH$_3$ SAM surface (Figure 2a).

To further reveal the relation between the observed slip behavior and microscopic characteristics of the water/solid interface, the water structure near SAM is characterized by calculating the water number density as a function of distance to the wall and normalizing it against the density at the middle of the channel (Figure 3a-c). The water density profile is further overlaid with the density profile of the SAM molecules. The water number density profiles exhibit oscillatory features near all the substrate walls, which is a clear indication of molecular layering. For all cases, the water density approaches bulk value when the distance from the SAM molecules is larger than ~ 0.4 nm. It is seen that as the interfacial adhesion increases from the hydrophobic -CH$_3$ to the hydrophilic -COOH surface, the first density peak becomes more pronounced near the wall. For all cases, an obvious valley appears after the first peak. This leads to a discontinuity within water and thus the slip can happen at the valley (~ 0.1-0.2 nm from the SAM) when the strong hydrophilic interfacial adhesion arrests the first layer. This agrees reasonably with the location of the no-slip boundary (~ 0.1 nm) that can be identified in Figure 2c. If we further zoom in the water/SAM interface region (Figure 3d-f), we can see that when SAM becomes more hydrophilic, there is much larger of water molecular density overlapping with the SAM density profile, indicating more significant water interdigitation. The calculated number of interdigitated water molecules in the SAM molecules show 10, 95, 208 (out of the total 4000 water molecules) trapped inside the SAM molecules for the three surfaces, respectively. We also observe that the water molecules penetrated into the SAM layer are not permanently trapped in it but can move in and out even for the most hydrophilic -COOH surface. The enhanced water interdigitation phenomena is the result of the stronger SAM-water interaction which can impede the mobility of water molecules near the interface and thus lead to larger friction and thus the no-slip boundary condition.

With the water density profile and the velocity profile calculated, we can study the boundary slip velocity for each interface at each shear rate. This slip velocity is calculate as the difference between the imposed shearing velocity of the Au substrate and the mean flow velocity of the interfacial water layer, which is defined by the location of the first peak of the density profile of water.[47] Figure 4 shows the calculated results for the three surfaces as a function of shear rate. As shown in the figure, there is a linear relation between the slip velocity and the nominal shear rate. This relationship falls into the linear Navier boundary condition, $v_s = L_s \dot{\gamma}$, where $L_s$ is the slip length and $\dot{\gamma}$ is the imposed shear rate, i.e., the slope of the linear relation in Figure 4 is the slip length according to the Navier boundary condition. The calculated Navier slip lengths are 0.54 nm, 0.36 nm, 0.19 nm for -CH₃, -OH, -COOH, respectively, whereas the fitted slip lengths from Figure 2 are 0.90 nm, 0.05 nm, -0.10 nm. Both the traditional Navier slip length and the extrapolated slip length from Figure 2 are shear rate-independent, but what the traditional Navier boundary assumption fails to capture is the transition from the positive slip to the negative slip phenomena. This is understandable since the definition of the boundary slip velocity used to obtain the Navier slip length makes it always positive. However, for the -COOH functionalized surface, there is actually no molecular slip as shown in the velocity profile in Figure 2c, where the slip boundary is displaced into the water.

Since the slip phenomenon is still mostly in the linear Navier boundary condition (Figure 4), we further calculate the Navier shear viscosity according to:[48]

$$\eta = -\frac{S_{xz}}{\dot{\gamma}} \qquad (2)$$

where the $S_{xz}$ is the shear component (xz) of the stress tensor, which consists of the kinetic energy contribution and the virial term.[49,50] The calculated shear viscosity values under different shear rates are shown in Figure 5a. Our calculated water viscosity is on the same order of magnitude as, but smaller than, the bulk water shear viscosity ($0.88 \; mPa \cdot S^{-1}$) under un-confined conditions.[51] This finding is consistent with other previous studies on the shear viscosity of nano-confined water.[52–54] In our cases, from the hydrophilic surface to the hydrophobic surface, the shear viscosity increases. This should be related to the stronger interfacial water-wall interaction at the hydrophilic surfaces which makes the liquid denser near the interface. As the shear velocity increases, the shear viscosity decreases for each surface functionalization (known as shear thinning[55]), which means that the increase of the shear stress in water does not grow proportionally as the increasing nominal shear rate according to the Navier shear viscosity equation (Eq. 2). Another parameter to illustrate the effect of the interfacial interaction is the friction coefficient, which can be calculated according to the Amontons-Coulomb law: $f = \frac{F_L}{F_N}$.[13] Here, $F_L$ and $F_N$ are the block-averaged lateral force and the normal force between the SAM-functionalized substrate and the water molecules calculated during shearing, respectively. As shown in Figure 5b, the friction coefficient increases monotonically as the increase of the shear velocity for the -OH and -COOH surfaces. For the -CH$_3$ SAM surface, higher shear velocity cannot lead to larger friction coefficient anymore due to the weak interaction between water and the -CH$_3$ SAM. More importantly, it is obvious that as the surface becomes more hydrophilic, the friction coefficient becomes greater.

Higher local viscosity and larger friction coefficient both lead to slowed water dynamics near the surface and thus impede water slip. We calculate the residence time of interfacial water molecules by evaluating the survival time correlation function $C_R(t)$ as:

$$C_R(t) = \frac{1}{N_w} \sum_{i=1}^{N_w} \frac{\langle P_{Rj}(0) P_{Rj}(t) \rangle}{\langle P_{Rj}(0) \rangle^2} \tag{3}$$

where $P_{Rj}$ is a binary function that equals to 1 if the $j$th water molecule resides in the interfacial water region for a time duration of $t$. As shown in Figure 6, clearly, the interfacial water residence time depends on different types of SAM layers. The interfacial water near the -CH$_3$ SAM surface decays much faster than the -OH and -COOH SAM surfaces. We use an exponential function to extract the decay time constants[56,57] for the three surfaces and found values of 22.9 ps, 29.5 ps, and 35.2 ps, respectively. The longer decay time indicates the slower migration of the interfacial water molecules on the SAM surface.

All the above observations should be rooted from the interfacial water-wall interaction. We further calculate the water-wall interaction potential energy as a function of space in the three different functionalized nanochannels. The potential energy experienced by the water molecules is calculated by summing up the van der Waals and the electrostatic energies of the water molecules interacting with the Au/SAM substrates (Figure 7). We find that the interfacial water-wall potential energy averaged over a distance of 1.2 nm (20% greater than the force cutoff range) from the wall increases from 0.1 to 0.3 and then to 0.4 kcal/mol as the surface changes from hydrophobic to hydrophilic functionalization. The stronger interfacial interaction can attract more water molecules to the surface to make the interfacial liquid layer denser[35] (as indicated by the density profiles in Figure 3). These results explain the observed trend in shear viscosity and friction coefficient at the water-wall interfaces as surface functionalization changes.

**CONCLUSION**

In summary, we have performed MD simulations to study the slip behavior of water confined between different SAM-functionalized surfaces. We observe a positive-to-negative slip transition from hydrophobic to hydrophilic SAM functionalizations, which is found related to the stronger interfacial interaction between water molecules and more hydrophilic SAM molecules. The stronger interaction increases the surface friction and local viscosity, making water slip more difficult. More hydrophilic functionalization also slows down the interfacial water relaxation and leads to more pronounced water trapping inside the SAM layer, which both impede water slip. The results from this work show that using SAM functionalization can be a practical and effective means to control the water slip in nanochannels. This work will also provide useful insights into the understanding of the water slip at functionalized surfaces and design guidelines for various applications.

**ACKNOWLEDGEMENTS**

D.H. acknowledges the financial support from the Chinese Scholarship Council. T.Z., E.L. and T.L. thank the support from the NSF (1706039), the Center for the Advancement of Science in Space (GA-2018-268), and the Dorini Family for the endowed professorship in Energy Studies. The simulations are supported by the Notre Dame Center for Research Computing, and NSF through XSEDE computing resources provided by SDSC Comet and Comet and TACC Stampede under grant number TG-CTS100078. L.X. would like to appreciate the support from Research Incentive Performance Program of Chongqing Science and Technology Bureau (cstc2018jxjl20004). Z.Q. would like to thank the support from the Basic Science Center Program for Ordered Energy Conversion of the National Natural Science Foundation of China (No.51888103).

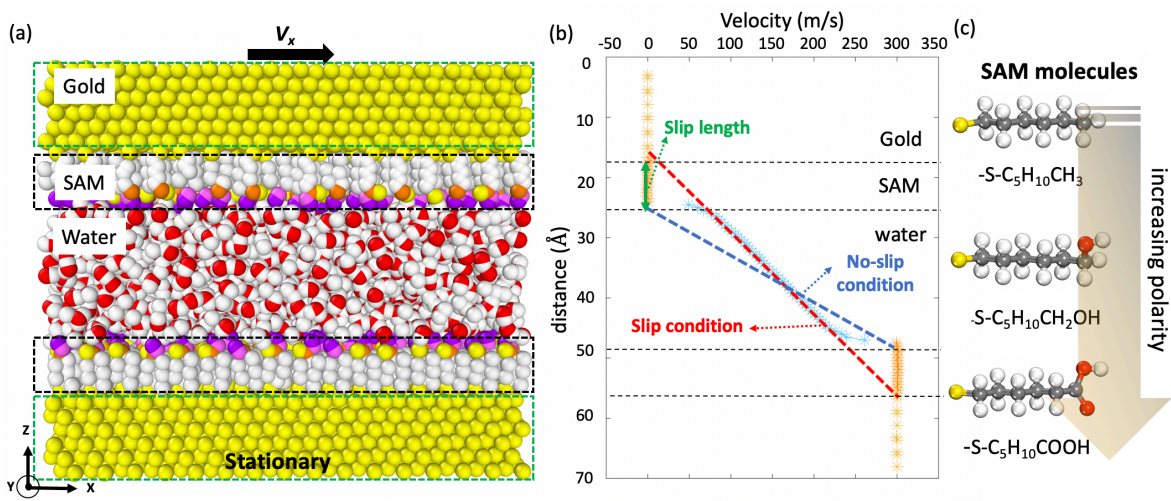

Figure 1. (a) An example model setup for shearing simulations, where the SAM layers are thermostatted at 300 K during the shearing process to prevent heating. The top gold substrate moves at different speed in the x-direction. (b) Example velocity profile showing how the slip length is extrapolated in NEMD simulations. The blue dashed line shows the no slip velocity profile. The red dashed line shows the slip velocity. The green double-headed arrow between the red and blue dashed lines shows the fitted slip length. (c) Chemical structures of SAMs with different polarities studied in this work.

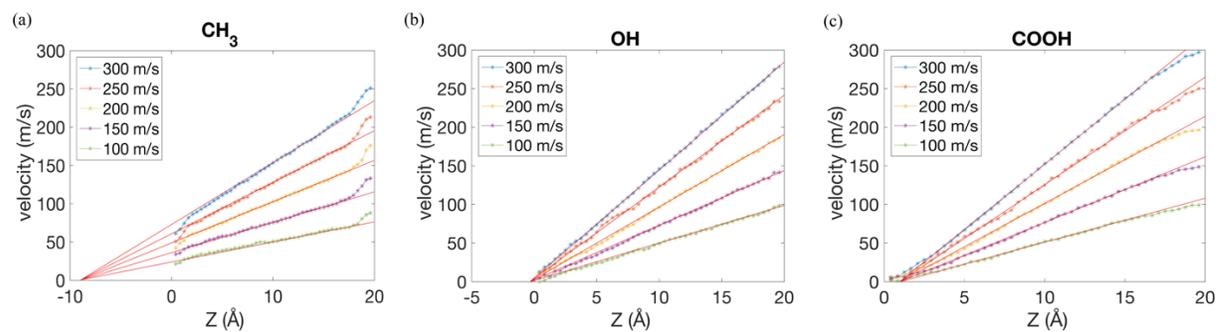

Figure 2. TIP3P Water velocity profile for (a) -CH$_3$ SAM, (b) -OH SAM and (c) -COOH SAM systems under different shear velocity. The red dotted lines are the fitting of the linear portion of each velocity profile.

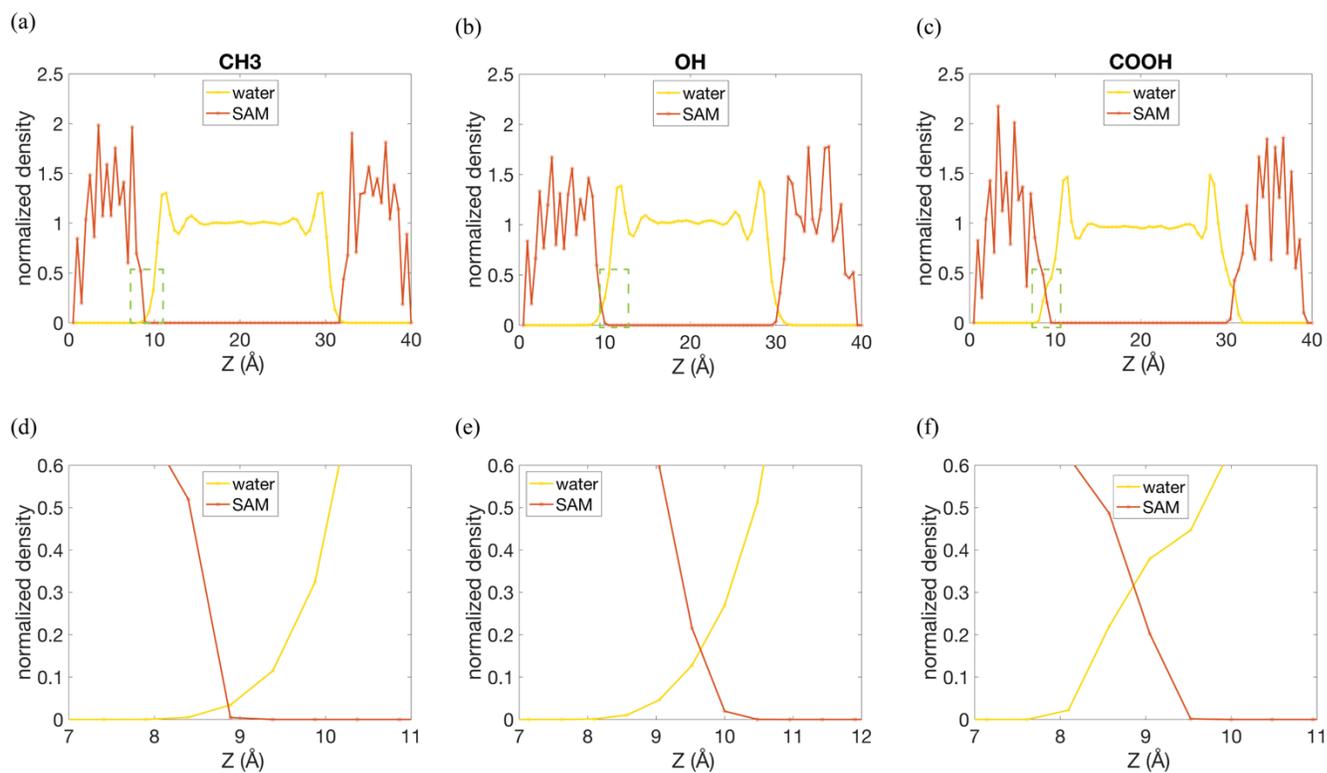

Figure 3. (a-c) The normalized density profile of water and SAM for the -CH₃ SAM, -OH SAM, and -COOH SAM surfaces. The dashed-green box is the zoomed-in regions, which are shown in (d-e) to illustrate the SAM-water interdigitation for the three surfaces.

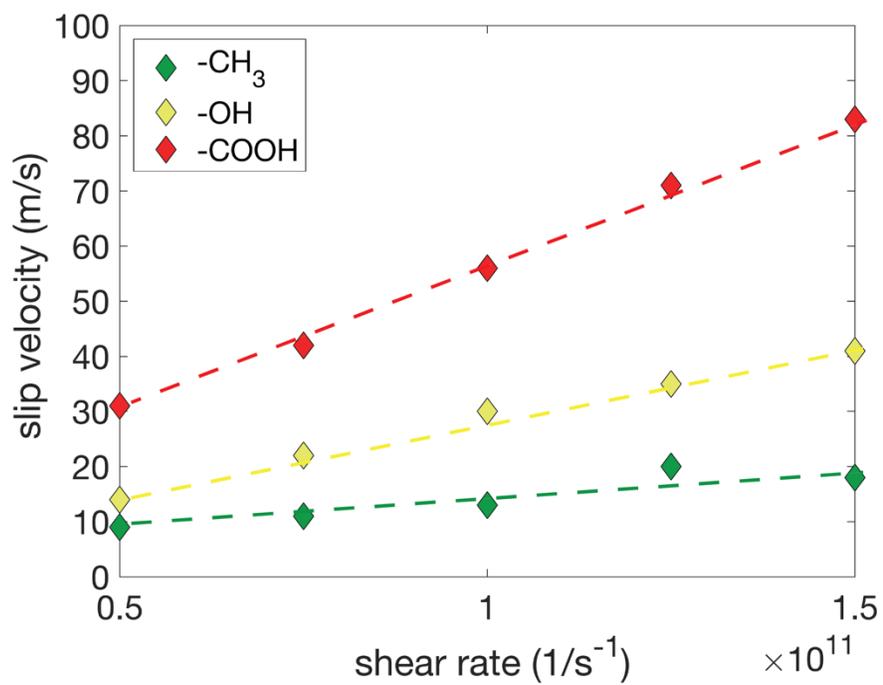

Figure 4. Boundary slip velocity of the confined water as a function of the shear rate in three types of SAM functionalized substrate.

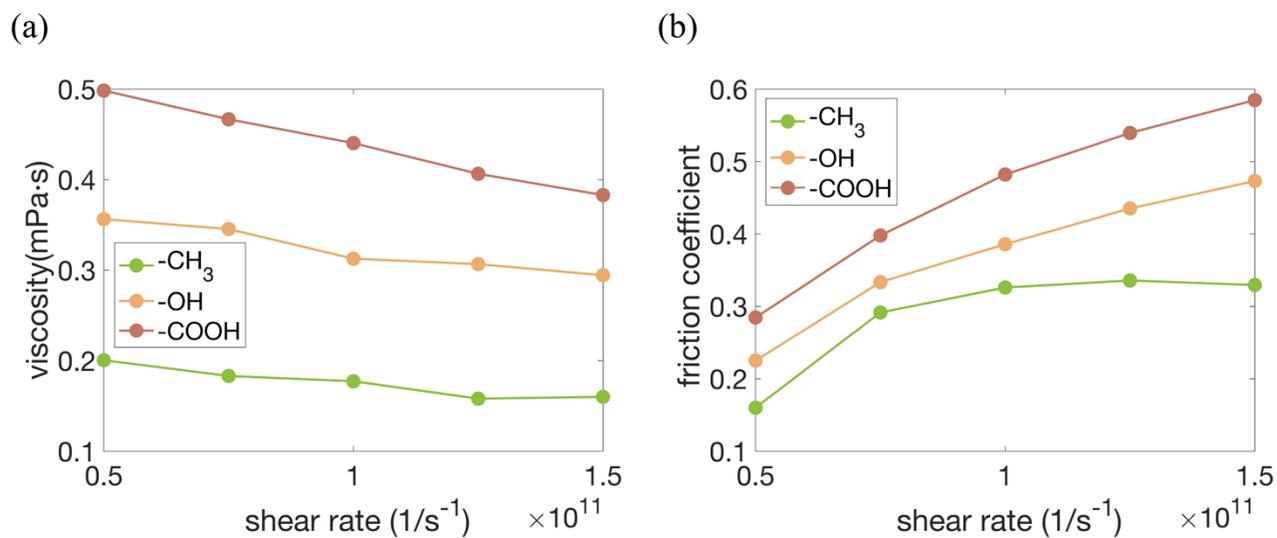

Figure 5. (a) The shear viscosity of water confined between the walls, and (b) the friction coefficient as a function of the shear rate for the three different SAM-functionalized surfaces.

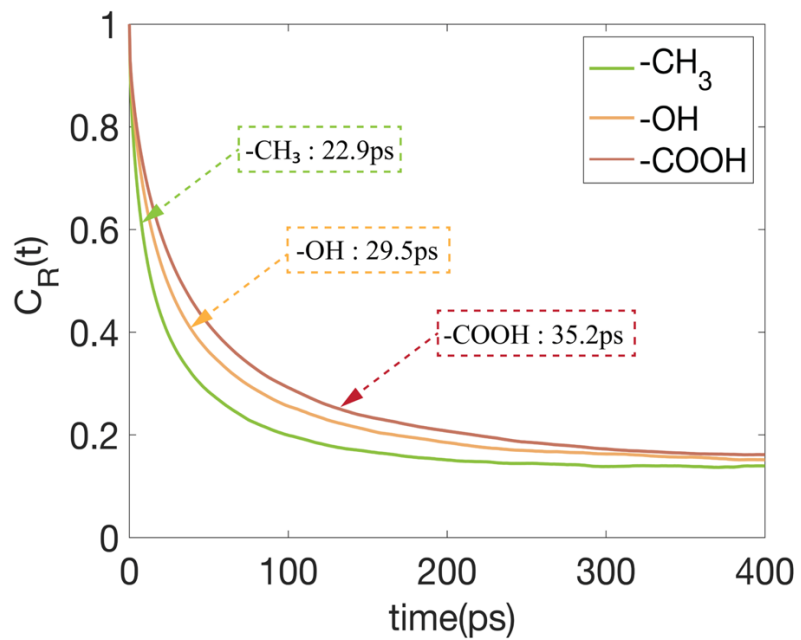

Figure 6. The residence time of interfacial water molecules near the -CH₃, -OH, and -COOH SAM-functionalized surfaces.

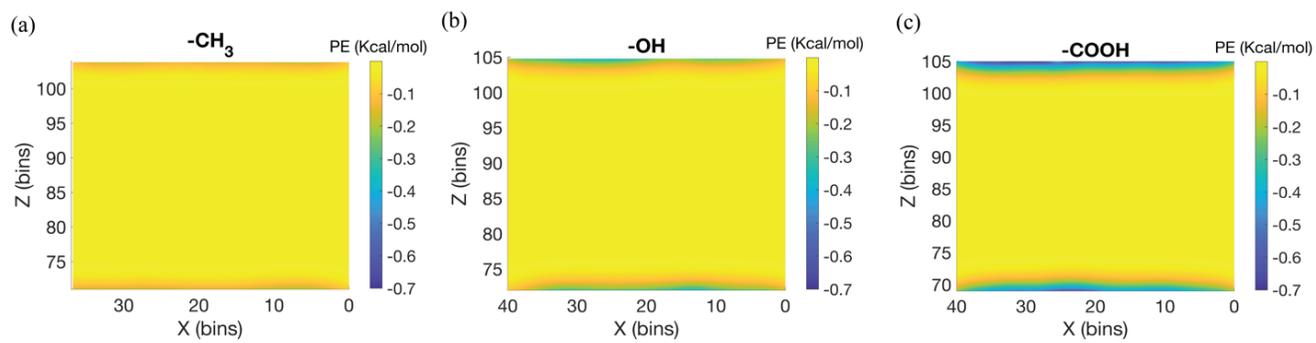

Figure 7. Potential energy profile in the x-z plane experienced by water confined between the three different functionalized Au substrates.

## Supporting Information

**Tuning Water Slip Behavior in Nanochannels using Self-Assembled Monolayers**

Dezhao Huang[1], Teng Zhang[1], Guoping Xiong[2], Linji Xu[3], Zhiguo Qu[4,*], Eungkyu Lee[1,*],

Tengfei Luo[1,5,*]

1.  Department of Aerospace and Mechanical Engineering, University of Notre Dame, Notre Dame, Indiana 46556, USA

2.  Department of Mechanical Engineering, University of Nevada – Reno, Reno, Nevada 89557, USA

3.  Chongqing Research Academe of Environmental Science, Chongqing 401120, China

4.  Moe Key Laboratory of Thermo-Fluid Science and Engineering, Energy and Power Engineering School, Xi'an Jiaotong University, Xi'an 710049, China

5.  Department of Chemical and Biomolecular Engineering, University of Notre Dame, Notre Dame, Indiana 46556, USA

*Corresponding authors:  Email: tluo@nd.edu, elee18@nd.edu, zgqu@mail.xjtu.edu.cn

# 1. SPC/E water velocity profiles

To make sure our slip findings is not model-dependent, we also perform simulations using the SPC/E water model for all different shear velocity cases. The positive to negative slip transition is still observed for the SPC/E water model as shown in Fig. S1.

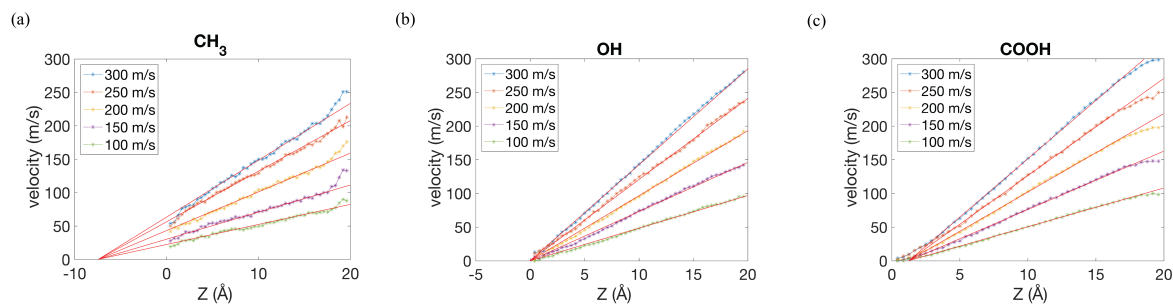

**Figure S1**. SPC/E water velocity profile for (a) -CH₃ SAM, (b) -OH SAM and (c) -COOH SAM systems under different shear velocity for SPC/E water model. The red dotted lines are the fitting of the linear portion of each velocity profile.

# 2. Force Field Parameters

The parameters used in this study for TIP3P[1] and SPC/E[2] water model are listed in Table S1.

**Table S1. Parameters for SPC/E and TIP3P water molecular models**.

| Model | $\sigma$ (Å) | $\varepsilon$(kJ/mol) | l(Å) | $q_H$(e) | $q_O$(e) | $\theta_0$ |
|-------|------|-----------|------|--------|--------|------|
| SPC/E | 3.166 | 0.65 | 1.0000 | +0.4238 | -0.8476 | 109.47 |
| TIP3P | 3.15061 | 0.6364 | 0.9572 | +0.4170 | -0.8340 | 104.52 |